# THE JOINT MILLI-ARCSECOND PATHFINDER SURVEY (JMAPS): MISSION OVERVIEW AND ATTITUDE SENSING APPLICATIONS

Bryan N. Dorland,[*] Rachel P. Dudik,[†] Zachary Dugan,[‡] and Gregory S. Hennessy[§]

The Joint Milliarcsecond Pathfinder Survey (JMAPS) is a Department of Navy bright star astrometric all-sky survey scheduled for launch in the 2012 timeframe. Mission objectives include a complete update of star positions for the 2015 epoch to accuracy levels of 1 milliarcsecond (5 nano-radians) for bright stars, as well as demonstration of 10 milliarcsecond attitude determination capability and 50 milli-arcsecond attitude control on-orbit. In the following paper, we describe the general instrument design and expected performance. We also discuss the new mission capabilities enabled by the unprecedented attitude determination accuracy of such an instrument, and focus specifically on the application to long distance (50,000-100,00 km) formation flying and solar system navigation.

## INTRODUCTION

The Joint Milli-Arcsecond Pathfinder Survey (JMAPS) is a Department of the Navy (DoN) space astrometry mission, approved for flight, with a 2012 launch date. JMAPS is an all-sky, bright-star astrometric and spectrophotometric survey. The primary goal of the mission is to completely update the bright star catalogs currently used by Department of Defense (DoD), NASA and civilian sensors for purposes of attitude determination. Secondary goals include the development and flight of cutting-edge hardware that will benefit future attitude sensing and imaging applications. In addition, the instrumentation developed to collect stellar catalog data will also demonstrate unprecedented attitude determination capabilities, useful to future advanced applications.

JMAPS is currently under development, with the program office at the Office of Naval Research (ONR), the Principal Investigator and ground data processing activity at the US Naval Observatory (USNO), and the space, downlink and mission operations, and launch segment activity at the Naval Research Laboratory (NRL). The current concept, shown in Figure 1, is of a single-aperture instrument hosted on a microsat spacecraft bus. The instrument is similar in concept and size to a star tracker, but with significantly higher accuracy.

---

[*] Chief, Astrometric Satellite Division, Astrometry Department, U.S. Naval Observatory, 3450 Mass. Ave. NW, Wash. DC 20392.
[†] Astronomer, Astrometry Department, U.S. Naval Observatory, 3450 Mass. Ave. NW, Wash. DC 20392.
[‡] Astronomer, Astrometry Department, U.S. Naval Observatory, 3450 Mass. Ave. NW, Wash. DC 20392.
[§] Astronomer, Astrometry Department, U.S. Naval Observatory, 3450 Mass. Ave. NW, Wash. DC 20392.

The instrument observes the sky in a step-stare mode, spending approximately thirty seconds (includes integration, slew and settle) on each star field. Target stars within the magnitude range of 0.5—14 will be observed 50 to 75 times over the course of three years. These data are returned to the ground, where they are processed together with instrument state parameters, thereby yielding a "global solution" for each star; i.e., a solution for the five primary astrometric parameters (position in Right Ascension [RA] and declination [DEC], proper motion in RA and DEC, and parallax). The baseline orbit for the satellite is a 900 km Sun Synchronous Orbit (SSO).

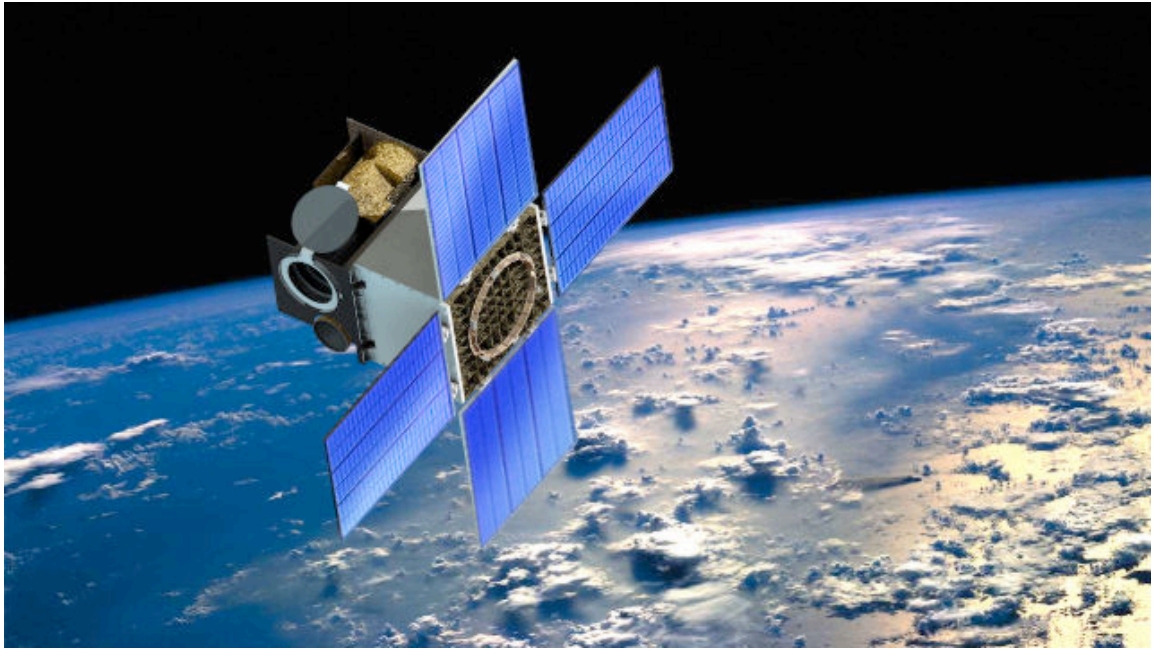

**Figure 1. Current JMAPS spacecraft concept. Primary instrument is an 8" visible/near IR telescope. Spacecraft is microsat-class.**

The mission will produce a final catalog by the end of 2016 that will include stars within the target magnitude range at mission accuracies of 1 milli-arcsecond (mas)[*]/1 mas year$^{-1}$/1 mas for position/proper motion/parallax. In addition, the catalog will include photometric and spectral data that will support extending high-accuracy astrometric and photometric results across the visible/near IR spectral range.

In this paper, we will first describe the on-orbit instrument and the attitude sensing capabilities of the JMAPS telescope. We will then provide illustrations of how these new capabilities can be used to support future missions by discussing two specific applications: (1) long distance formation flying, and (2) solar system navigation.

---

[*] 1 mas is approximately equal to 5 nano-radians.

**THE JMAPS INSTRUMENT**

The heart of the JMAPS program is the astrometric instrument. The instrument consists of a single-aperture[*] optical telescope assembly (OTA), the Focal Plane Assembly (FPA) and the supporting electronics. The telescope is an 8", on-axis astrograph design, implemented using silicon carbide and designed to be highly stable over the on-orbit conditions. The focal plane consists of an 8k x 8k, 10 um pixel CMOS-Hybrid FPA. The FPA is controlled by the instrument electronics, which also digitizes, processes and stores the output data from the FPA for periodic download to the ground station. The total instrumental spectral response in the astrometric passband is approximately aligned with the astronomical photometric Cousin's I-band[†]. The instrument is deployed on a "deck" as shown in Figure 2. In the figure, the OTA, the camera electronics and the coarse star tracker (used during acquisition and lost-in-space modes) are shown. The instrument deck is mounted to the spacecraft during integration.

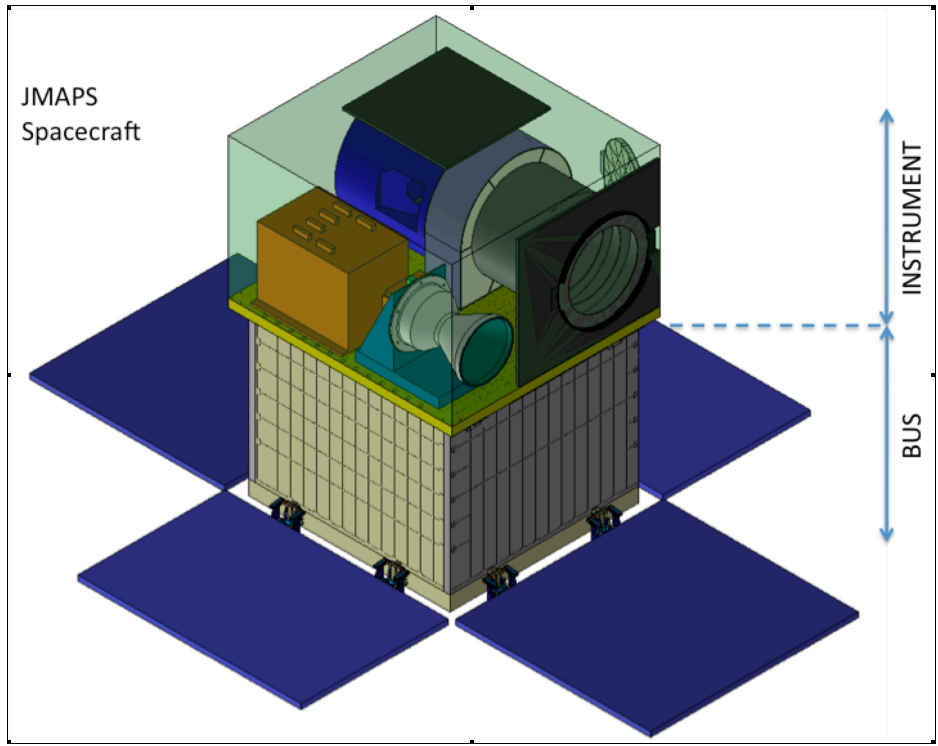

**Figure 2. JMAPS spacecraft concept showing separation of bus and instrument sections. On the instrument deck, the primary telescope is shown, along with the electronics (orange box) and the secondary, coarse star tracker. The flat plate at the aperture is the dedicated FPA radiator.**

---

[*] The basis for the single-aperture, step-stare imager for global astrometry, a significant departure from the two-aperture approach adopted by other space astrometry missions such as Hipparcos and Gaia, was established by Zacharias & Dorland (reference 1).
[†] Approximately 700—900 nm

The total single-measurement systematic floor of the instrument (i.e., residual FPA, electronics and optical effects) has been designed to not exceed 5 mas. For a 12$^{th}$ magnitude star in the primary astrometric band, combining the signal-to-noise centroiding accuracy with the predicted systematic floor yields a single measurement precision of at least 7 mas for a twenty-second integration time, the longest integration time routinely used for the survey. By combining multiple observations over the three-year mission lifetime, final catalog accuracies of 1 mas or better will be achieved.

**ATTITUDE SENSING AND POINTING CONTROL**

During science data collection, the spacecraft is required to maintain very stable pointing. In order to achieve the single-axis pointing requirement of 50 mas (1-$\sigma$), the spacecraft Attitude Determination and Control System (ADCS) will use the astrometric instrument as the fine guidance sensor. The astrometric instrument will observe approximately 12 bright stars per field of view, and read out star images every 200 msec. The instrument will then calculate individual centroids for these twelve guide stars, combine them into a single solution for the boresight, and generate a boresight quaterion at a 5 Hz rate.

Using these individual star position measurements, how well can the instrument boresight pointing be determined? Analysis of USNO's NOMAD catalog statistics (Reference 2) suggests that the mean guide star will have an I-band magnitude of 8—9. Adopting the more conservative value of 9$^{th}$ magnitude, the current instrument model predicts a single measurement precision (including both random and systematic error) of less than 13 mas. Combining all twelve of these measurements to determine the overall boresight orientation of the instrument, pointing accuracies of well under 10 mas are feasible with significant margin. Sparse fields, such as those near the galactic poles, yield the worst-case results. In these cases, NOMAD statistics suggest that the guide star population can be as faint at 11$^{th}$ magnitude. This translates to single measurement precisions of approximately 32 mas per star per 200 msec integration time. Here too, total boresight accuracies of under 10 mas are feasible, though with significantly less margin than the average case. This analysis is consistent with more detailed analyses conducted by mission ADCS personnel as part of ongoing assessment and design activities.

**TWO APPLICATIONS: LONG-DISTANCE FORMATION FLYING AND SOLAR SYSTEM NAVIGATION**

In order to illustrate the value of this level of attitude determination for future space missions, in the following sections we provide two illustrations of how a JMAPS-like instrument could be used to enable new in-flight capabilities for advanced missions. In the first section, we examine whether a JMAPS-like instrument can be used to enable long-distance formation flying and what accuracies can be achieved. In the second section, we look at the ability of a JMAPS-class instrument to determine position in the solar system, for support in spacecraft navigation when outside the Earth orbital environment.

**Long-distance formation flying**

One potential future application for this class of instrument is to help in the alignment of components of formation flying systems. In Figure 3, we consider the alignment of two components (marked "A" and "B") that are separated by 50,000 km. In this particular example, the alignment tolerances are extremely tight in the transverse direction, but relatively loose in the radial direction.

In order to accomplish this alignment, a JMAPS-derived astrometric instrument is deployed onto A and pointed at B. On B, an I-band beacon is included, so that A can assume a steady level of illumination from B rather than counting on reflectance of solar illumination, which can fluctuate with the varying solar zenith angle. We assume that the beacon has sufficient power to appear 12th magnitude when viewed from A. This assumption will be discussed in more detail later in the paper.

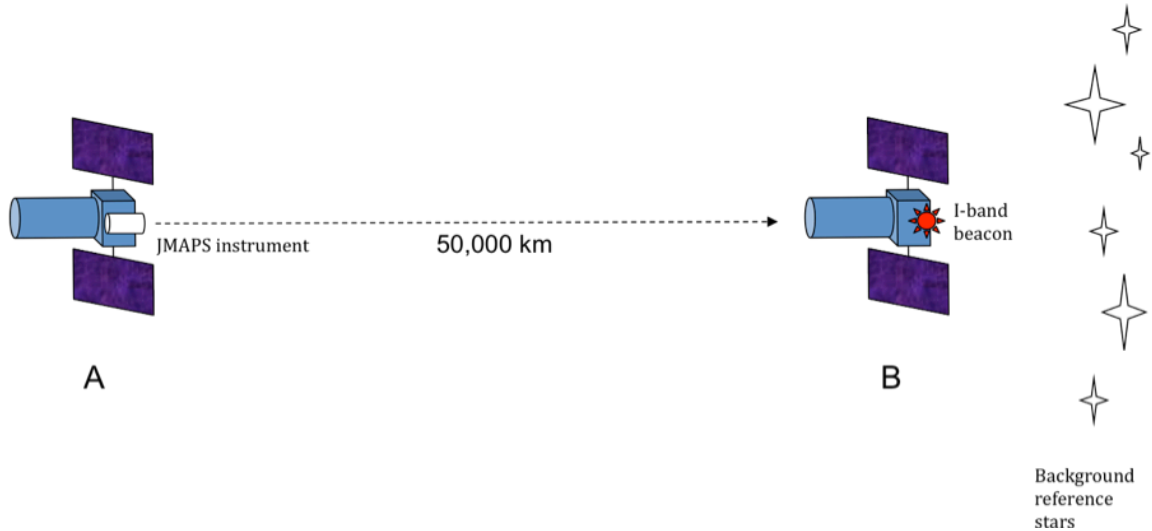

**Figure 3. Formation flying alignment problem. Two spacecraft, marked "A" and "B" in diagram, must be aligned to very small tolerances in the transverse direction. An astrometric instrument is deployed on A and a beacon on B. Alignment is effected using the instrument to guide A into alignment with B. Based on analysis described in the text, alignment accuracies of 2 meters in the transverse direction are feasible.**

The JMAPS instrument will be able to determine its boresight orientation to better than 10 mas every 200 msec. For a 10 second integration time, a 12th magnitude I-band beacon can be detected and measured to approximately 8 mas accuracy. Ten seconds is equivalent to 50 measurement cycles for guide stars. By combining guide star observations over ten seconds, the instrument can reduce boresight orientation error to under 1 mas. The combined accuracy of the observations of the beacon against the background reference grid is approximately 8 mas, which translates to 2 m at 50,000 km. This means that using a JMAPS-class instrument and a beacon, at a minimum, the two spacecraft can be aligned to within 2 meters at a distance of 50,000 km.

Is the beacon requirement feasible? Analysis of the flux at the instrument aperture indicates that a 4.5 kW source is needed on B to produce a 12th magnitude source at A if the beacon uniformly illuminates $4\pi$ sr. On the other hand, a unidirectional beacon reduces the power requirement on the source significantly. A directional beam of width* of approximately 11° from B to-wards A would reduce power requirements to a feasible 10 W power level. Such a beam width corresponds to approximately 10,000 km at 50,000 km distance, well within the navigational capabilities of A and B even without using the JMAPS-class instrument. Thus we conclude that 2m precision in formation flying can be achieved at 50,000 km given a 10 W beacon directed toward a JMAPS-class instrument.

---

* i.e., beam diameter

**Solar system navigation**

In this application we consider solar system navigation. Around the Earth, spacecraft typically use Global Positioning System (GPS) signals to locate their positions and orbits with high accuracy. However, outside of low Earth orbit, such positioning is impossible to any accuracy. For spacecraft deployed to Sun-Earth L2—an increasingly popular destination for astronomical missions—accurate position determination can be an extremely challenging problem. Figure 4 illustrates the way in which an astrometric instrument can solve this problem.

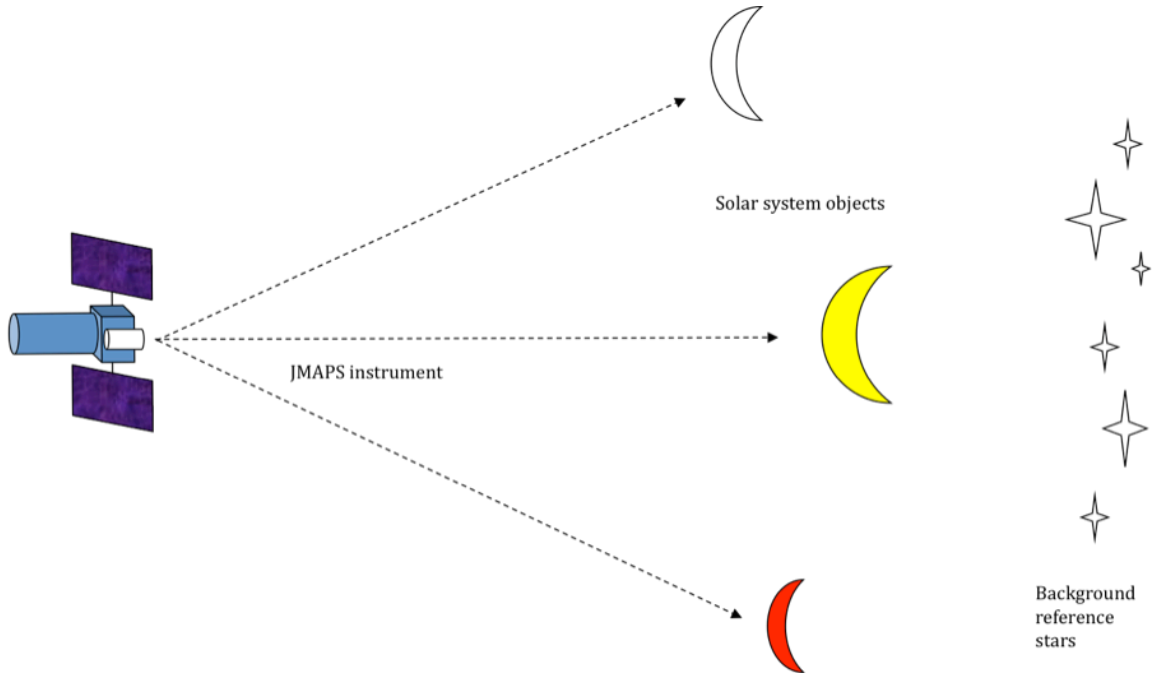

**Figure 4. Solar system navigation problem. 6 DOF Spacecraft position and velocity needs to be determined with some accuracy. Spacecraft uses JMAPS-like astrometric instrument to determine solar system position.**

Given solar system reference objects with accurately known positions, Figure 4 illustrates how observations from a JMAPS-class instrument can be used to find the position of the spacecraft within the solar system. The astrometric instrument takes bearings on these objects against the background reference grid, and calculates the position of the observer using the known positions and motions of the solar system reference objects.

Choosing the appropriate set of reference objects is crucial to this approach. On the one hand, giant planets are bright and have well defined orbits. However these objects also present a number of obstacles including: (1) their paucity—in this case position determination accuracy becomes highly dependent on the geometry of the scenario, and (2) they are resolved—for a JMAPS-class instrument, the definition of centroid and center of mass are problematic for such objects. In addition, for resolved objects such as Jupiter and Saturn, the photocenter will a function of solar zenith angle (i.e., the illumination conditions) at the levels of accuracy we are concerned with.

A better solution employs a set of reference targets such as asteroids that are relatively numerous, more or less uniformly distributed around the Sun, and relatively point source-like. Just such a population exists—90 km-class asteroids. There are about 100 of these objects with relatively well-know orbits. These are main belt objects that are approximately evenly distributed around the sun. They are large enough to be quasi-spherical in nature, and as a result, photocenters fluctuations due to rotation are estimated to be at the 5 mas level (Reference 3). This is small enough to support the navigation accuracy needs.

Our method involves observing approximately 8 of these asteroids near opposition, approximately 8 positioned about a month ahead of the Earth's orbit and approximately 8 that are about a month behind for a total of 24 objects. The eight asteroids at opposition will have apparent magnitudes around 12 in the I-band, while the leading and trailing 16 will have apparent magnitudes of approximately 13 due to distance and illumination effects. For a ten-second integration time (as with the formation flying scenario above), these objects can be observed with sub-10 mas precision. By combining the accuracies over these 24 objects, we have derived the resultant error ellipsoid for our solar system position measurements. The error ellipsoids are shown in Figure 5.

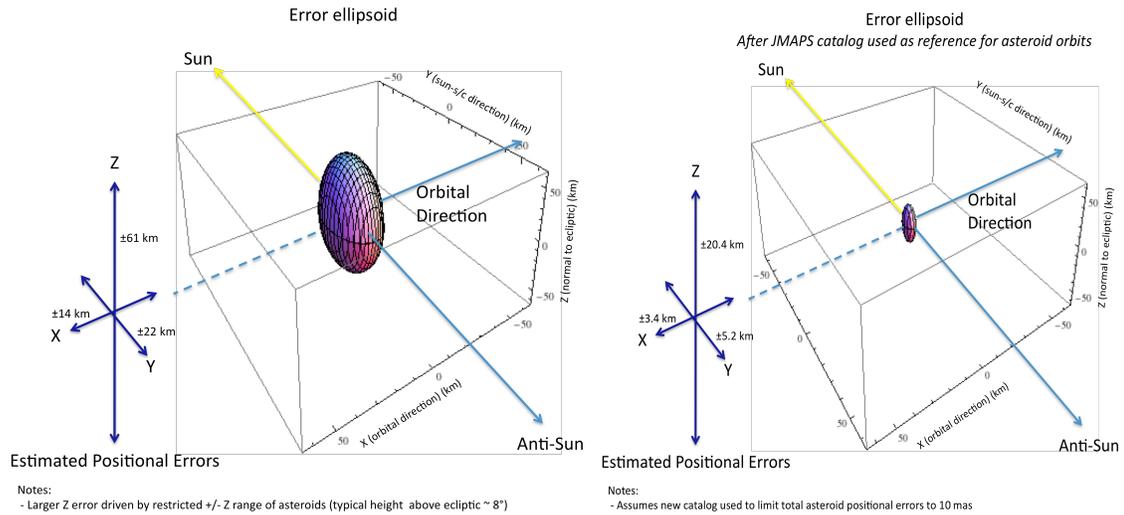

**Figure 5. Solar system navigation results. (Left) results assuming no improvements to background reference catalogs, and (right) results assuming updated background catalogs. The Z directed error is much larger than Z or Y because the asteroids typically orbit ~8° or less out of plane, so not much information is collected in the Z direction.**

In the first plot (left figure), the error ellipsoid associated with using the astrometric instrument and the current generation of catalogs is shown. Errors of order a couple dozen km are obtained in-plane, while the out-of-plane error is over 60 km. These errors are dominated by errors in positions of the asteroids caused by the zonal errors in the star catalogs. Significant improvement is obtained by using JMAPS-updated catalogs (right figure), which drive the star position errors down to 1 mas. Resultant observer platform position errors are reduced to a few kms in-plane, and approximately 20 kms out-of-plane.

## CONCLUSION

JMAPS is a DoN mission, scheduled for launch in 2012, that will update the bright star astrometry catalog. The mission will demonstrate new technological capabilities, including the capacity for an astrometric instrument to obtain very high precision pointing knowledge.

A JMAPS-class instrument can be used in advanced applications for complex missions as a star-tracker capable of unparalleled accuracy. We have discussed two examples of applications for this new and significantly improved attitude determination capability: long-distance formation flying and solar system navigation. We have shown that, using JMAPS-class astrometric instruments, alignment tolerances of 2 m at 50,000 km distances are feasible. We have also shown that, using the astrometric instrument to observe solar system reference objects—specifically, asteroids—one can determine position to a few km in-plane and a few tens of km out-of-plane.

## ACKNOWLEDGMENTS

The authors note that Webster Cash first suggested the solar system navigation application to us, and would like to acknowledge the assistance of Hugh Harris and George Kaplan in discussing these concepts.